# Magnetism, Charge Order and Giant Magnetoresistance in SrFeO$_{3-\delta}$ Single Crystals

A. Lebon[1], P. Adler[2], C. Bernhard[1], A. V. Boris[1], A. V. Pimenov[1], A. Maljuk[1], C. T. Lin[1], C. Ulrich[1] and B. Keimer[1]

[1] Max-Planck-Institut für Festkörperforschung,
Heisenbergstr. 1, D-70569 Stuttgart, Germany
[2] Institut für Anorganische Chemie
Engesserstraße, Geb.30.45
Universität Karlsruhe (TH), D-76128 Germany

(Dated: October 28, 2003)

## Abstract

The electronic and magnetic properties of SrFeO$_{3-\delta}$ single crystals with controlled oxygen content ($0 \leq \delta \leq 0.19$) have been studied systematically by susceptibility, transport and spectroscopic techniques. An intimate correlation between the spin-charge ordering and the electronic transport behavior is found. Giant negative as well as positive magnetoresistance are observed.

The relationship between spin ordering, charge ordering, and magneto-transport effects has recently received much attention in the context of the "colossal magnetoresistance" (CMR) phenomenon in manganites [1]. Experiments have revealed a key role of phase separation between charge ordered (CO) insulating and charge disordered metallic phases in driving the CMR effect [2], but many questions about the detailed interplay between spin, charge, orbital, and lattice degrees of freedom remain unresolved. A comparison to other compounds with similar electronic structures can provide insight into the physical properties of the manganites. Iron(IV) perovskites form a Ruddlesden-Popper series of lattice structures akin to the manganites and are particularly suitable for such a comparison, because the high spin $Fe^{4+}$ ion present in these phases [3] is isoelectronic to the Jahn-Teller ion $Mn^{3+}$: both ions have three electrons in the $t_{2g}$ level and one electron in the $e_g$ level of the crystal field generated by the surrounding oxygen octahedron. However, while the parent compound of the CMR manganites, $LaMnO_3$, is insulating and shows a cooperative Jahn-Teller effect, its analogue $SrFeO_3$ remains metallic and cubic at all temperatures [4,5]. Its helical magnetic structure [6,7] has no counterpart in the manganites, which generally exhibit collinear magnetic order [8]. The different electronic properties of manganites and ferrates have been ascribed to the more pronounced hybridization between transition metal and oxygen orbitals in the ferrates [9], but a detailed theoretical description is still lacking. Moreover, in contrast to the $Mn^{3+}$-$Mn^{4+}$ alternation observed in charge-ordered manganites, $SrFeO_3$ frequently adopts charge-disproportionated $Fe^{3+}$-$Fe^{5+}$ configurations when the valence electron density is modified by substitutions on the Sr site [3,10-12].

Most remarkably, recent experiments on polycrystalline samples of pseudocubic SrFeO$_{2.95}$ have uncovered anomalies in the magnetic susceptibility and electrical resistivity curves as well as a pronounced negative magnetoresistance (MR) below 50 K [13]. Substantial negative MR was also observed in Co-doped iron (IV) perovskites [14]. The origin of these effects is not yet clear, and substantial sample-to-sample variations were observed. In order to advance our understanding of the magneto-transport properties of the ferrates and their relation to magnetic and charge order, we have synthesized a series of SrFeO$_{3-\delta}$ single crystals with controlled oxygen content ($\delta$ < 0.2). It is known that the structural phase diagram of SrFeO$_{3-\delta}$ encompasses the stoichiometric cubic perovskite phase (C, $\delta$ =0) as well as the oxygen vacancy ordered tetragonal (T, $\delta$ = 0.125, Sr$_8$Fe$_8$O$_{23}$) and orthorhombic (O, $\delta$ = 0.25, Sr$_4$Fe$_4$O$_{11}$) phases separated by miscibility gaps [15, 16]. Our extensive investigations on SrFeO$_{3-\delta}$ single crystals with a variety of techniques including magnetometry, resistivity, Mössbauer spectroscopy and optical spectroscopy reveal a detailed picture of the evolution of the physical properties with oxygen concentration. Specifically, we report three different MR effects in this system. A large negative MR is observed for cubic SrFeO$_3$ near 55 K in the temperature range of a previously unobserved magnetic phase transition. A giant negative MR effect of up to 90% at a magnetic field of 9 T in mainly tetragonal SrFeO$_{2.85}$ is related to a combined Fe$^{3+}$-Fe$^{4+}$ charge - magnetic ordering transition near 70 K. Furthermore, a large positive MR is found in even more oxygen-deficient SrFeO$_{2.81}$ at low temperatures.

The experiments were performed on high quality SrFeO$_{3-\delta}$ single crystals synthesized in a floating zone furnace [17]. The oxygen contents were determined to an accuracy of ~0.02 by thermogravimetry in a reducing atmosphere. As-grown oxygen deficient crystals with δ = 0.13, 0.15, and 0.19 were obtained under slightly different growth conditions. Stoichiometric SrFeO$_3$ was prepared by post-annealing under high oxygen pressure of 5 kbars at 400°C. Mössbauer spectra (see below) confirmed that the high-pressure annealed sample consists of the pure C phase (δ = 0), whereas the δ = 0.13 and δ = 0.19 crystals are phase-separated with fractions of about 80% T / 20% C and 70% T / 30% O phase, respectively [18]. Powder x-ray diffraction patterns are consistent with the Mössbauer results. Magnetic susceptibility and electrical resistivity were measured with a PPMS system (Quantum Design). Cr/Au electrodes were deposited to perform four-point resistivity measurements. 57-Iron Mössbauer spectra on ground crystals were collected with a standard spectrometer operating with a sine-type drive signal and a $^{57}$Co source in a Rh matrix. All spectra are referenced to α-Fe. The infrared spectra were collected on an infrared ellipsometer at the ANKA synchrotron [19].

Fig. 1 shows the magnetic susceptibility of selected single crystals measured at 1 T in field cooling and heating runs. In the δ = 0 sample, a maximum heralding the onset of helical magnetic order at 130 K is observed together with a small cusp of the susceptibility at ~ 55 K. The latter shows a thermal hysteresis. The δ=0.15 sample exhibits a sharp maximum at 70 K with a noticeable thermal hysteresis and evidence of residual C phase with its characteristic peak at 130 K. The 70 K feature remains for δ=0.19.

We now turn to the magneto-transport properties of the SrFeO$_{3-\delta}$ system. In agreement with earlier work [5], the temperature dependence of the resistivity indicates that the C phase is metallic (Fig. 2a). In addition, however, a decrease in resistivity by a factor of 2 is observed at 55K. This transition exhibits a hysteresis and is obviously related to the 55 K anomaly in the susceptibility data. Application of a 9 T magnetic field shifts the transition to *higher* temperatures, and accordingly a large negative MR effect occurs in a narrow temperature range around 55 K ($\Delta T_{1/2} \sim 8K$, max. MR ~25%). The two oxygen deficient samples display a weakly activated, semiconducting behavior upon cooling from room temperature to 70 K, where the resistivity increases abruptly by an order of magnitude (Figs. 2b,c). The transition is characteristic of the T phase and shifted to *lower* temperatures by a magnetic field of 9 T, in contrast to the 55 K transition in cubic $SrFeO_3$. Although the two transitions are evidently of different origin, the field-induced shift of the transition again gives rise to a large negative MR effect that peaks sharply in a narrow temperature range. In particular, for $SrFeO_{2.85}$ a giant negative MR of 90% at ~69 K is found ($\Delta T_{1/2} \sim 3.8K$). Finally, a pronounced *positive* MR effect is apparent in the low temperature insulating phase of $SrFeO_{2.81}$.

In order to elucidate the nature of the magnetic phase transitions, we have carried out Mössbauer experiments on $\delta = 0$ and $\delta = 0.13$ samples. The spectra of the $\delta = 0$ sample below ~ 130 K (Fig. 3a) consist of a single six-line pattern due to magnetic hyperfine interactions (isomer shift IS = 0.16 mm s$^{-1}$, magnetic hyperfine field $B_{hf}$ = 32.3 T at 8 K) which verifies that only $Fe^{4+}$ is present. The shapes of the spectra are very similar below and above 55 K where the magnetization and resistivity data indicate an additional phase transition. No anomalies are observed in the temperature dependence of IS and $B_{hf}$. This excludes any change in the charge states of the iron ions and suggests that the 55 K transition corresponds to a rearrangement of the $Fe^{4+}$ moments [20].

In the spectra of the $SrFeO_{2.87}$ sample, on the other hand, there is clear evidence for a combined magnetic-CO transition. Above 130 K the whole material is in a paramagnetic state, and similar to earlier work on polycrystalline $SrFeO_{2.86}$ [21] the spectra (not shown) were analyzed in terms of two components only: one $Fe^{4+}$ single line and one average charge $Fe^{3.5+}$ quadrupole doublet. The charge assignments are based on the IS values (Fig. 3d). A first magnetic transition below ~130 K leads to single sextets D (Fig. 3b) that can be assigned to the cubic fraction of the sample by comparison to $SrFeO_3$.

The second transition, seen at 70 K in the magnetization data, leads to complicated but well-structured low-temperature spectra, which were successfully modeled by a superposition of five (15K) or six (>15K) hyperfine sextets. A typical decomposition is illustrated in Fig. 3c for 30 K, and the results of the data evaluation are given in Fig. 3d. From $B_{hf}(T)$ it is evident that the magnetic order for all components except D vanishes near 70 K. Thus all sites except D are considered as intrinsic to the T phase, which is the majority component in this sample. From the IS and $B_{hf}$ values, charge states of 3+, 4+ and 3.5+ are assigned to the sites (A, B), (D, E, F) and C, respectively. Considering the low-temperature area fractions and assuming equal Debye-Waller factors for the various Fe sites, this leads to an average iron oxidation state of 3.74 and an oxygen content of 2.87 ($\delta = 0.13$), in good agreement with the thermogravimetric analysis. The area fraction of the $Fe^{3.5+}$ sites decreases from ~55% in the paramagnetic phase to ~10% in the magnetically ordered phase. This suggests that the magnetic ordering of the T phase near 70 K coincides with charge ordering of $Fe^{3.5+}$ into $Fe^{3+}$ and $Fe^{4+}$. The sum of the area fractions of the $Fe^{3+}$ sites A and B is approximately equal to the area fraction of the $Fe^{4+}$ site F, which supports the CO scenario [22]. The residual $Fe^{3.5+}$ sites in the CO state could be related to an oxygen deficiency in the present T phase compared to stoichiometric $Sr_8Fe_8O_{23}$. A charge disproportionation of $Fe^{4+}$ is excluded as this should lead to higher $Fe^{3+}$ fractions, as for instance observed in $Sr_{2/3}La_{1/3}FeO_{3-\delta}$ [23].

Additional support for a CO transition is obtained from infrared ellipsometry measurements on the $\delta = 0.15$ sample. Fig. 4 shows that the optical conductivity suddenly drops between 90 and 70 K, in agreement with the dc resistivity (Fig. 2b). Further, a change of the lattice structure related to the CO process is indicated by the appearance of numerous additional optical phonon modes at low temperatures. The real part of the dielectric function, also displayed in Fig. 4, confirms the Kramers-Kronig consistency of all phonon features. Studies of the low-temperature crystal structure of the T phase are required before an assignment of these modes can be obtained.

In summary, we have observed three different magnetoresistance effects in $SrFeO_{3-\delta}$ single crystals ($0 \leq \delta \leq 0.19$). Two of these effects are superficially similar, but arise from entirely different physical mechanisms: First, the large negative MR in cubic $SrFeO_3$ does not involve charge order and is probably related to a rearrangement of the helical structure of $Fe^{4+}$ moments. Second, a combined $Fe^{4+}$ - $Fe^{3+}$ charge – magnetic ordering, which is different from the $Fe^{3+}$-$Fe^{5+}$ CO in other iron(IV) containing ferrates [10-12], leads to an even larger negative MR in oxygen-deficient samples containing the T phase as majority component. The giant negative MR near the CO transition of the T phase is reminiscent of a similar (though smaller) effect near the first-order Verwey transition in $Fe_3O_4$ [24]. Finally, the positive MR in $SrFeO_{2.81}$ is more difficult to understand. Positive MR due to a field-induced decrease of the localization length is commonly observed in the variable range-hopping regime of doped semiconductors, but these effects are typically smaller and are obtained for substantially higher fields [25].

Further experiments are required to ascertain whether an "extraordinary magnetoresistance" effect [26] due to small inclusions of the cubic metallic phase (below the sensitivity limit of the Mössbauer data) can account for this observation. In view of the present results it appears likely that the physical properties previously reported for oxygen deficient $SrFeO_{2.95}$ [13] reflect a superposition of effects due to the majority C phase and minority T phase. In contrast to Ref. [13], our $SrFeO_{3-\delta}$ single crystals do not show negative MR behavior below 50 K. The latter possibly corresponds to grain boundary effects in polycrystalline samples.

We thank S. Ahlert, E. Brücher, K. Förderreuther, W. Hölle, Y.-L. Mathis and F. Schartner for generous technical assistance, and L. Alff for valuable discussions.

[1] R. von Helmolt *et al.*, Phys. Rev. Lett. **71**, 2331 (1993).

[2] M. Uehara *et al.*, Nature **399**, 560 (1999).

[3] A. E. Bocquet *et al.*, Phys. Rev. B **45**, 1561 (1992).

[4] P. K. Gallagher *et al.*, J. Chem. Phys. **41**, 2429 (1964).

[5] J. B. Mac Chesney *et al.*, J. Chem. Phys. **43**, 1907 (1965).

[6] T. Takeda *et al.*, J. Phys. Soc. Jpn. **33**, 967 (1972).

[7] H. Oda *et al.*, J. Phys. Soc. Jpn. **42**, 101 (1977).

[8] E. O. Wollan and W.C. Koehler, Phys. Rev. B **100**, 545 (1955).

[9] J. Matsuno *et al.*, Phys. Rev. B **60**, 4605 (1999).

[10] J. Q. Li *et al.*, Phys. Rev. Lett. **79**, 297 (1997).

[11] S. K. Park *et al.*, Phys. Rev. B **60**, 10788 (1999).

[12] T. Ishikawa *et al.*, Phys. Rev. B **58**, R13326 (1998).

[13] Y. M. Zhao *et al.*, Phys. Rev. B **64**, 024414 (2001).

[14] P. D. Battle *et al.*, Chem. Commun. 987 (1998); S. Ghosh and P. Adler, Solid State Commun. **116**, 585 (2000); P. Adler and S. Ghosh, Solid State Sci. **5**, 445 (2003).

[15] Y. Takeda *et al.*, J. Solid State Chem. **63**, 237 (1986).

[16] J. P. Hodges *et al.*, J. Solid State Chem. **151**, 190 (2000).

[17] A. Maljuk *et al.*, J. Cryst. Growth **257**, 427 (2003).

[18] The $\delta = 0.13(2)$ and $\delta = 0.15(2)$ samples have very similar susceptibility curves and thus similar phase compositions are assumed.

[19] C. Bernhard *et al.*, Thin Solid Films, in press (2003).

[20] In cubic SrFeO$_3$ the hyperfine field is determined entirely by the isotropic Fermi contact interaction. The Mössbauer spectra are therefore not sensitive to a possible spin reorientation.

[21] M. Takano *et al.*, J. Solid State Chem. **73**, 140 (1988).

[22] The two other Fe$^{4+}$ sites D and E correspond to the Fe$^{4+}$ fraction in the paramagnetic phase. Referring to the room temperature crystal structure of tetragonal Sr$_8$Fe$_8$O$_{23}$, the Fe$^{4+}$ signals in the paramagnetic phase were attributed to Fe(1)O$_5$ square pyramids and nearly regular Fe(3)O$_6$ octahedra whereas the Fe$^{3.5+}$ signals were assigned to the more distorted Fe(2)O$_6$ octahedra [16]. Component E in the low temperature spectra corresponds very likely to square pyramidal Fe$^{4+}$ sites, whereas sextet D comprises Fe$^{4+}$ sites from the C phase and from nearly regular FeO$_6$ octahedra in the T phase. This is in accord with the observed decrease in the area fraction of D between 70 and 90 K. The low-temperature spectra of the δ = 0.13 sample do not reveal any indications for the O phase Sr$_4$Fe$_4$O$_{11}$ which should lead to paramagnetic sites even at 15 K [21]. The latter are indeed observed in the spectra of sample δ = 0.19.

[23] P. D. Battle *et al.*, J. Solid State Chem. **77**, 124 (1988).

[24] R. Rosenbaum, Phys. Rev. B **63**, 094426 (2001).

[25] V. V. Gridin *et al.*, Phys. Rev. B **53**, 15518 (1998).

[26] S.A. Solin *et al.*, Science **289**, 1530 (2000).

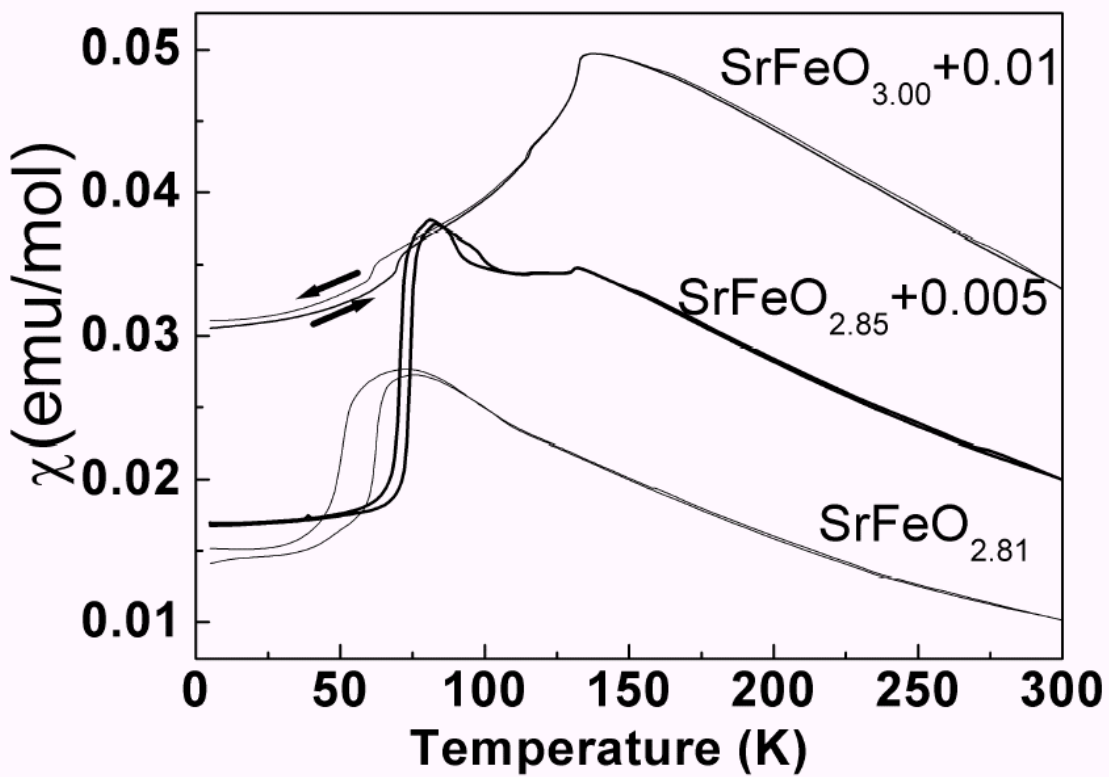

Fig. 1. Susceptibility of SrFeO$_{3.00}$, SrFeO$_{2.85}$ and SrFeO$_{2.81}$ single crystals, measured in field cooling and subsequent field heating runs at B=1T. The curves for the latter two samples were shifted in $\chi$ by the amounts indicated in the legend.

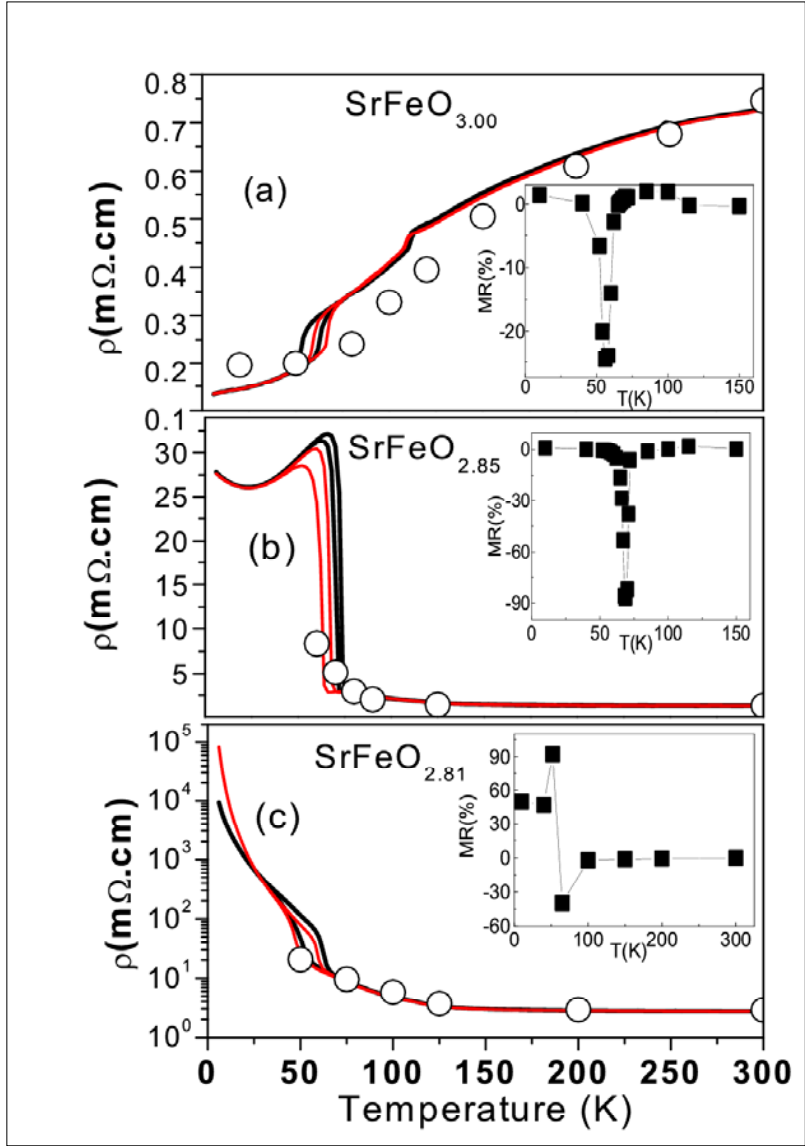

Fig. 2. Resistivity in zero field cooling and heating runs (black) compared to field cooling and heating runs with a 9 Tesla field (red) for a) $SrFeO_{3.00}$, b) $SrFeO_{2.85}$ and c) $SrFeO_{2.81}$. As the absolute value of the resistance measurements (lines) was influenced by microcracks, these data were normalized around room temperature to IR data extrapolated to zero frequency (open circles; see Fig. 4 for IR data on $SrFeO_{2.85}$). Insets: Magnetoresistance (MR) measured at 9 Tesla after isothermal field scans between 0 and 9T. MR is defined as $(\rho(H)-\rho(0))/\rho(0)$.

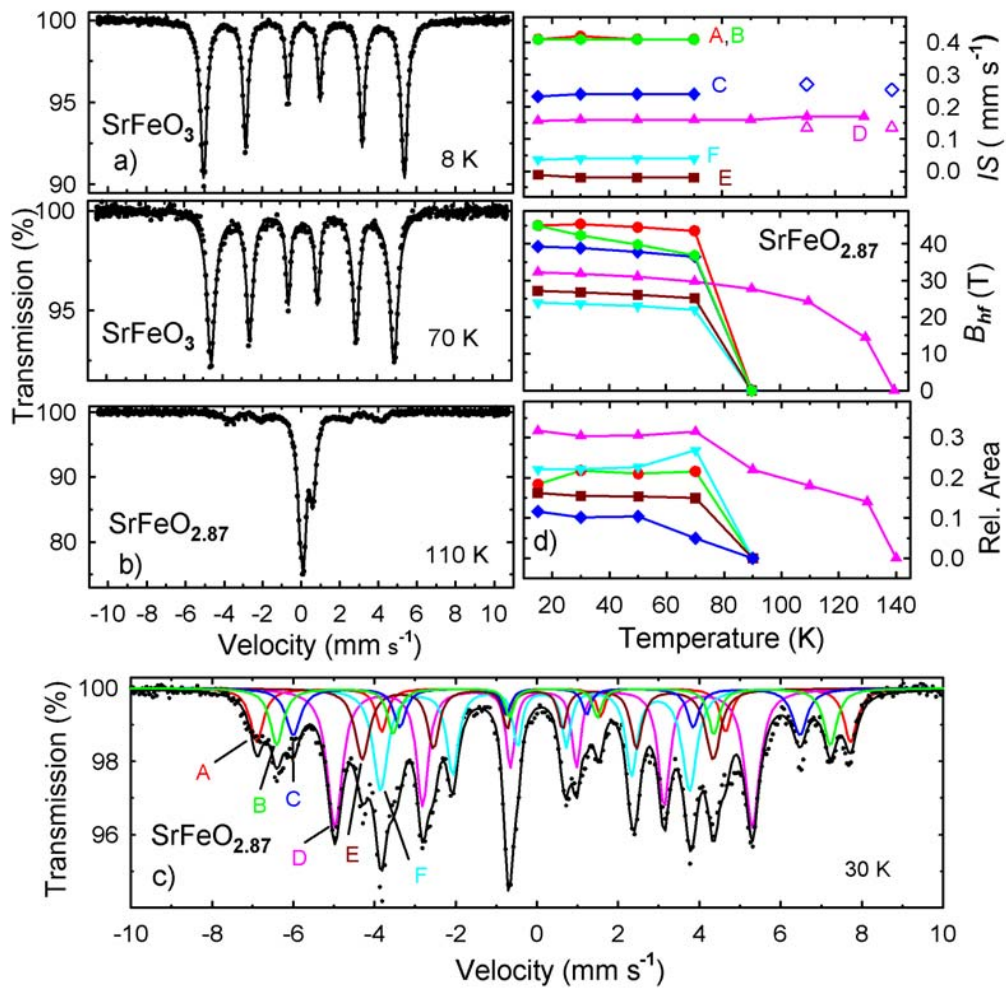

Fig. 3. a) Selected Mössbauer spectra of (a) cubic $SrFeO_3$ and (b) mostly tetragonal $SrFeO_{2.87}$. c) Decomposition of the 30 K spectrum of $SrFeO_{2.87}$ into sub-spectra. d) Temperature dependence of isomer shifts IS (top), hyperfine fields $B_{hf}$ (middle), and area fractions (bottom) for the various iron sites of $SrFeO_{2.87}$. Full and open symbols correspond to the magnetically ordered and paramagnetic phases, respectively. In the area fraction plot the full circles correspond to the sum of sites A and B.

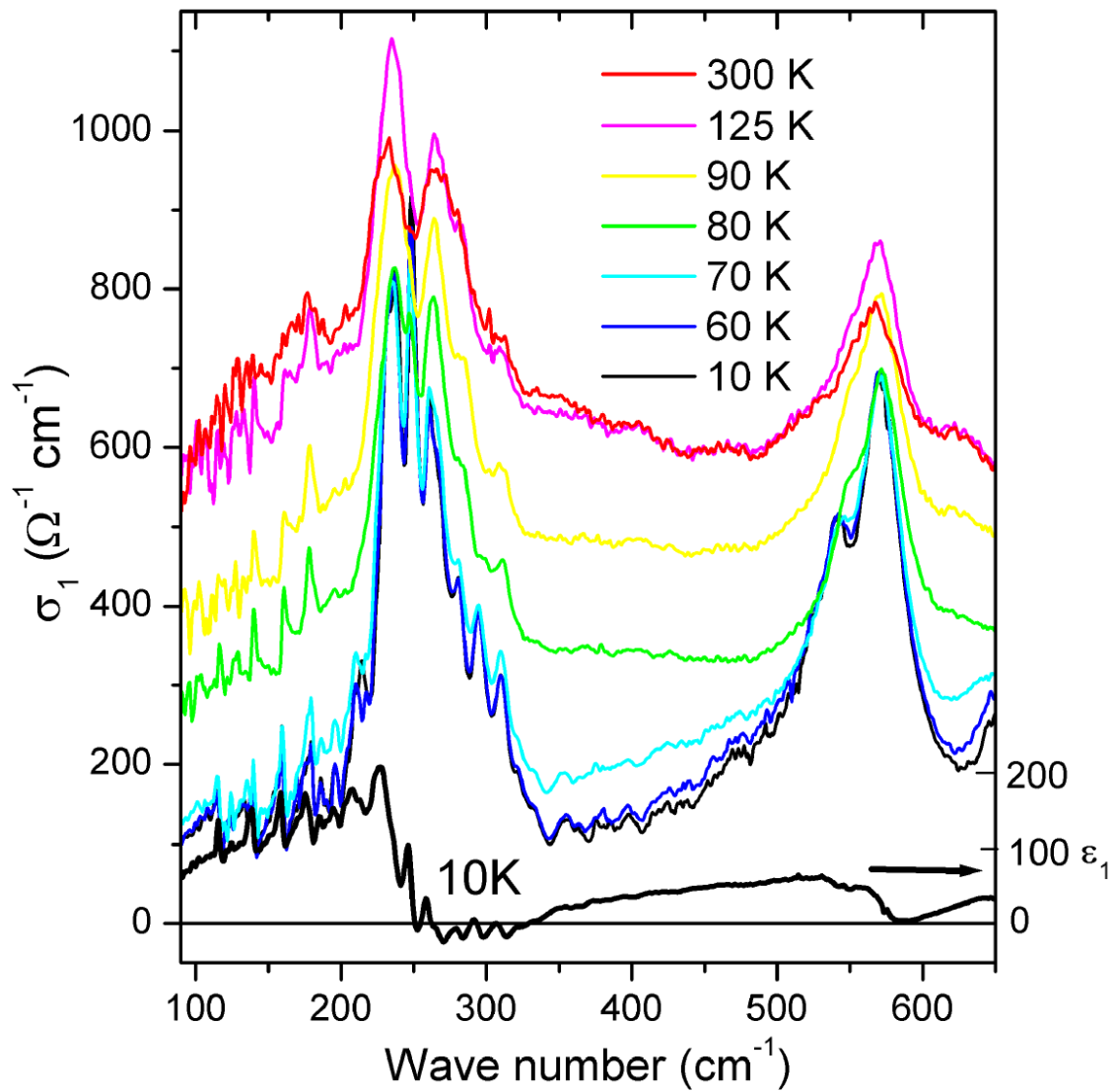

Fig. 4. Real part of the IR conductivity $\sigma_1(\omega)$ of $SrFeO_{2.85}$ measured at different temperatures. The bottom line shows the real part of the dielectric function $\varepsilon_1(\omega)$ at T = 10 K.